\documentclass[aps,prl,twocolumn, superscriptaddress,nofootinbib,groupedaddress]{revtex4}  

\usepackage{graphicx}
\usepackage{epsfig}
\usepackage{bm}
\usepackage{amsmath}
\usepackage{amssymb}
\usepackage{wasysym}
\usepackage{epstopdf}
\usepackage{color}
\usepackage{xcolor}
\usepackage{epstopdf}
\usepackage{upgreek}
\usepackage{arydshln}
\usepackage[T1]{fontenc}
\usepackage{natbib}

\usepackage{psfrag}
\usepackage{color}
\usepackage{graphicx,float,amssymb}

\usepackage{color}
\usepackage{graphicx}
\usepackage{bm}
\usepackage{amsmath}

\usepackage[colorlinks=false, bookmarks=true]{hyperref}

\newcommand{\etal}{\emph{et al.}}

\usepackage{graphicx}
\usepackage{epsfig}
\usepackage{bm}
\usepackage{amsmath}
\usepackage{amssymb}
\usepackage{wasysym}
\usepackage{epstopdf}
\usepackage{color}
\usepackage{xcolor}
\usepackage{epstopdf}
\usepackage{upgreek}
\usepackage{arydshln}
\usepackage[T1]{fontenc}

\begin{document}

\title{Dynamics and wakes of rising and falling cylinders}
\title{Continuous transition in the dynamics and wakes of rising and falling cylinders triggered by particle moment of inertia}

\title{Mass and moment of inertia govern the transition in the dynamics and wakes of freely rising and falling cylinders}
\author{Varghese Mathai}
\affiliation{ 
Physics of Fluids Group and Max Planck Center Twente, J. M. Burgers Centre for Fluid Dynamics, University of Twente, P.O. Box 217, 7500 AE Enschede, The Netherlands.}

\author{Xiaojue Zhu}
\affiliation{ 
Physics of Fluids Group and Max Planck Center Twente, J. M. Burgers Centre for Fluid Dynamics, University of Twente, P.O. Box 217, 7500 AE Enschede, The Netherlands.}

\author{Chao Sun}
\thanks{chaosun@tsinghua.edu.cn}
\affiliation{Center for Combustion Energy and Department of Thermal Engineering, 
Tsinghua University, 100084 Beijing, China.}
\affiliation{ 
Physics of Fluids Group and Max Planck Center Twente, J. M. Burgers Centre for Fluid Dynamics, University of Twente, P.O. Box 217, 7500 AE Enschede, The Netherlands.}

\author{Detlef Lohse}
\thanks{d.lohse@utwente.nl}
\affiliation{ 
Physics of Fluids Group and Max Planck Center Twente, J. M. Burgers Centre for Fluid Dynamics, University of Twente, P.O. Box 217, 7500 AE Enschede, The Netherlands.}
\affiliation{
Max Planck Institute for Dynamics and Self-Organization, 37077 G\"ottingen, Germany.}

\date{\today}

\date{\today}

\begin{abstract}
\textcolor{black}{In this \textit{Letter}, we study the motion and wake-patterns of freely rising and falling cylinders in quiescent fluid. 
We show that the amplitude of oscillation and the overall system-dynamics are intricately linked to two parameters: the particle's mass-density relative to the fluid $m^* \equiv \rho_p/\rho_f$  and its relative moment-of-inertia $I^* \equiv {I}_p/{I}_f$. This supersedes the current understanding that a critical mass density~($m^*\approx$ 0.54) alone triggers the sudden onset of vigorous vibrations. 
Using over 144 combinations of ${m}^*$ and $I^*$, we comprehensively map out the parameter space covering very heavy ($m^* > 10$) to very buoyant~($m^* < 0.1$) particles. The entire data collapses into two scaling regimes demarcated by a transitional Strouhal number, $St_t \approx 0.17$. $St_t$ separates a mass-dominated regime from a regime dominated by the particle's moment of inertia. A shift from one regime to the other also marks a gradual transition in the wake-shedding pattern: from the classical $\textit{2S}$~(\textit{2-Single}) vortex mode to a $\textit{2P}$~(\textit{2-Pairs}) vortex mode. \textcolor{black}{Thus, auto-rotation can have a significant influence on the trajectories and wakes of freely rising isotropic bodies.}
}
\end{abstract}

\maketitle

%
%
%
%
%
%
%
%

\textcolor{black}{Path-instabilities are a common observation in the dynamics of buoyant and heavy particles. Common examples are the fluttering of falling leaves and disks, and the cork-screw and spiral trajectories of air-bubbles rising in water~\cite{auguste2013falling,kelley1997path,ern2012wake}.  The oscillatory dynamics of such particles can vary a lot depending on the particle's size, shape and its inertia, and the surrounding flow properties. This can be important in a variety of fields ranging from sediment transport and fluidization to multiphase particle- and bubble-column reactors~\cite{richardson1954sedimentation,stringham1969behavior,hartman1993free}.}


{\color{black} Examples of organisms that exploit path-instabilities are plants and aquatic animals. These often make use of passive appendages attached to their bodies~(plumed seeds, barbs, tails, and protrusions) to generate locomotion~\cite{lacis2014passive,bagheri2012spontaneous,dickinson2000animals}.
Recently, L{\=a}cis \etal~\cite{lacis2014passive} demonstrated that the interaction between the wake of a falling bluff body and a protrusion clamped to its rear end can generate a sidewards drift by means of a symmetry-breaking instability similar to that of an inverted pendulum. Such kinds of passive interactions are advantageous to locomotion, since no energy needs to be spent by the animal.  Instead,  the energy can be extracted through fluid-structure interaction. 
}

\begin{figure}[!htbp]
	\centerline{\includegraphics[width = 0.5 \textwidth]{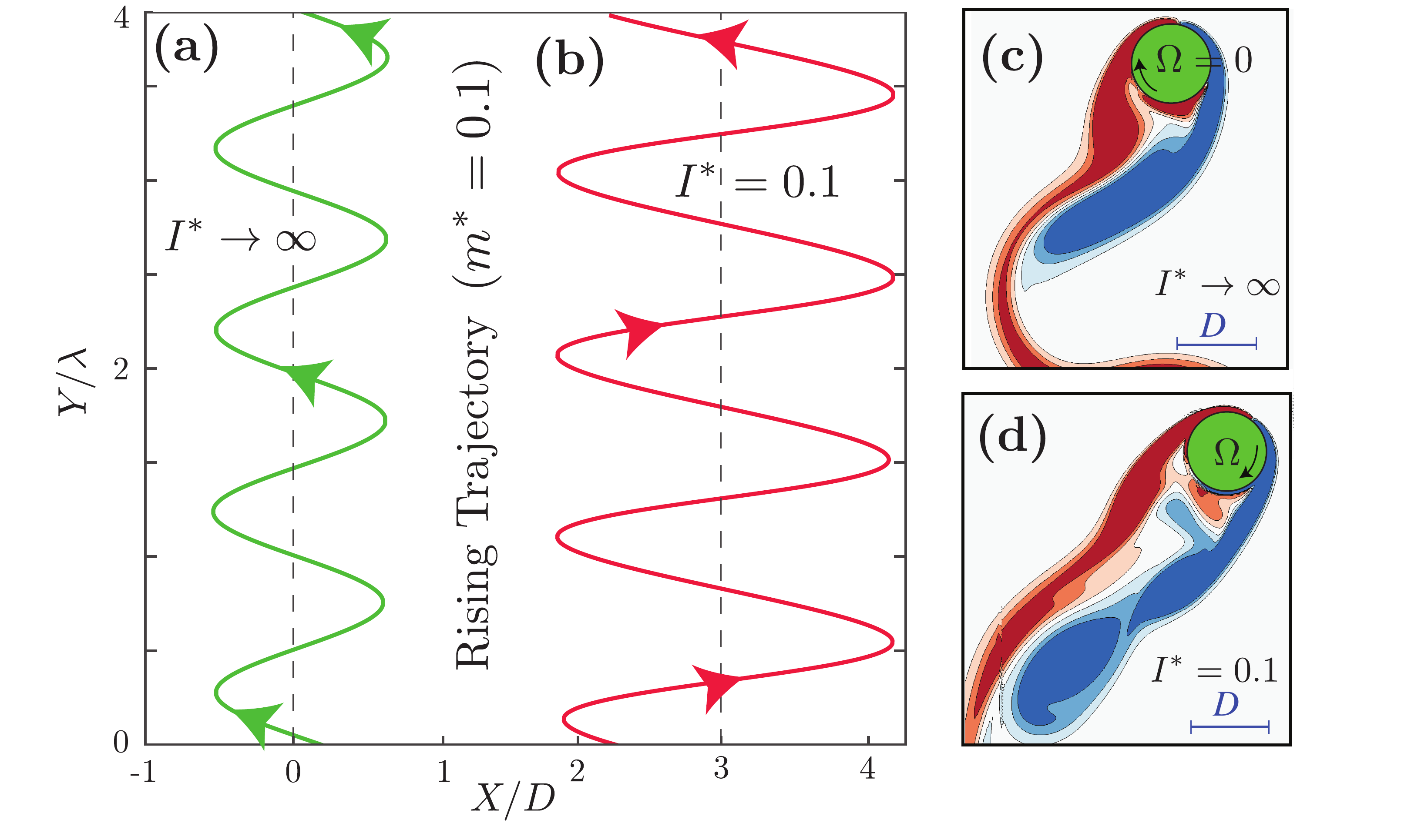}
	}
	\caption{\textcolor{black}{(a) Oscillatory rise trajectory of a buoyant cylinder~($m^* = 0.1$), with a constraint on rotation~($\Omega =0$). (b)~Oscillatory rise trajectory of the same cylinder with a low moment of inertia~($I^* = 0.1$). Here $\Omega$ is the angular velocity, $D$ is the diameter of the particle, and $\lambda$ is the wavelength of the oscillation. (c) \& (d)~Vorticity snapshots for the cylinders in (a) \& (b), respectively. Angular velocity $\Omega$ is zero for the rotationally constrained cylinder. Red-white-blue~colormap is used for the vorticity. 
	Allowing auto-rotation changes the wake pattern dramatically.}}
	\label{fig:trajectory_separation_circulation}
\end{figure}


\textcolor{black}{The simplest case of a rising or falling body in a fluid is a sphere or a cylinder released in quiescent fluid. This problem has traditionally been studied and characterized using two non-dimensional parameters:
the solid/fluid density-ratio (${m}^* \equiv \rho_p/\rho_f$) and the generalized particle Galileo number, $Ga$~\cite{jenny2004instabilities,veldhuis2004motion}. Among these, $Ga$ governs the onset of various kinds of wake-instabilities behind the particle, and ${m}^*$ governs the motion of the particle in response to these flow instabilities and vortex-induced forces. A number of investigators have studied the influence of these parameters, and various kinds of paths and wakes have been observed for rising and falling bodies~\cite{horowitz2006dynamics,horowitz2008critical,horowitz2010effect}.
However, despite a fair level of understanding of the mechanisms affecting path- and wake-instabilities, it remains unclear as to what factors precisely trigger vigorous path oscillations for a rising or falling object, such as a sphere or a cylinder.}

\textcolor{black}{The motion of a freely rising/falling particle  in a fluid is a complex fluid-structure-interaction~(FSI) problem. Wake-induced forces cause the particle to move, which in turn changes the flow field around it. This holds many similarities to the popular subject of vortex-induced-vibrations~(VIV)~\cite{govardhan2005vortex}. In this area, investigators have explored the dynamical response of elastically mounted and tethered bodies in uniform flows. A spring-mass-damper model was often used to make predictions of the oscillatory response of such systems under various conditions~\cite{horowitz2006dynamics,horowitz2008critical}. For example, a critical mass-ratio ${m}^*_{{crit}}$ was predicted for the sudden appearance of path-oscillations for elastically mounted spheres and cylinders undergoing VIV~\cite{govardhan2005vortex,williamson2004vortex}. For the freely rising body as well, similar predictions were made (${m}^*_{{crit}} = 0.54$ for cylinder, and ${m}^*_{{crit}} = 0.61$ for sphere) by modeling it as a spring-mass-damper system with zero spring-stiffness and zero damping~\cite{horowitz2010effect}. While a dependence on the mass-density ratio is undisputable, others observed wide scatter in their data to the point that there is no consensus with regard to whether a unique critical mass-ratio exists or not~\cite{karamanev1996dynamics,spelt1997motion}. Moreover, from a dynamical point of view, the existence of a $m^*_{crit}$ lies in contradiction with a fundamental concept: i.e, the motion of a light-particle in a fluid should be fluid added-mass dominated~(since ${m}_{a} > {m}_{p}$)~\cite{maxey1983equation,mathai2015wake}. Therefore, it remains surprising that a marginal reduction in mass-density alone would trigger the sudden appearance of large-amplitude oscillations, since the effective mass of the system (actual + added mass) is almost unchanged~\cite{ryskin1984numerical}. We will provide a plausible explanation for this anomaly.}

{\color{black} \textcolor{black}{In this {\it Letter}, we study the two-dimensional motion of circular cylinders rising or falling through a quiescent fluid. The body geometry~(circular cylinder), the state of the surrounding fluid~(quiescent), and the imposition of two-dimensionality of the flow make the problem simplified~(see Leontini \etal~\cite{leontini2007three} for a discussion on the possible three-dimensional effects). 
Nevertheless, this model problem can provide important clues about the underlying dynamics of buoyancy-driven bodies in general~\cite{namkoong2008numerical}. }We performed direct numerical simulations~(DNS) using the immersed boundary method. The solver uses the discrete stream-function formulation for the incompressible Navier-Stokes equations~\cite{zhu2014numerical} with a virtual force implementation~\cite{schwarz2015temporal}, which enables us to deal with both light and heavy particles. 
A rectangular computational domain of the size $100{D} \times 16D$ is used, with a grid width of $0.01L$ near the cylinder, where $D$ is the cylinder diameter.

	\begin{figure}[!htbp]
		\centerline{\includegraphics[width = 0.5\textwidth]{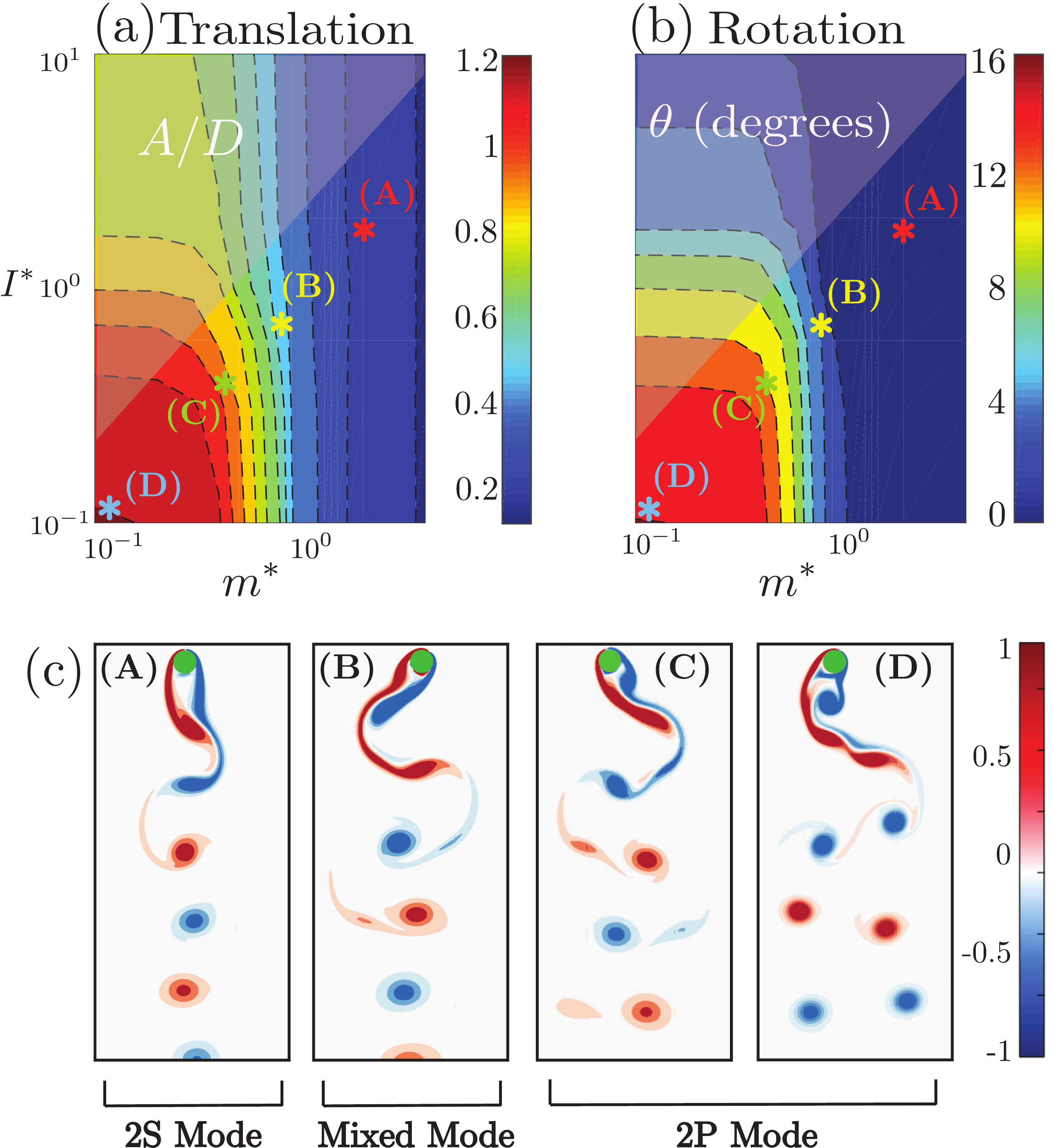}
		}
		\caption{\textcolor{black}{Contour map of the (a) translational and (b) rotational amplitudes of motion for buoyant~($m^* <1$) and heavy cylinders~($m^* >1$) in the [$m^*-\text{I}^*$] parameter space at Ga~$\approx$~500. The shaded regions mark the very high values of $I^*$, which are primarily of mathematical interest. (c)~Vorticity plots for the points marked as (A), (B), (C), and (D) in (a) and (b). 
		The vorticity is scaled to the [-1; 1] range. The corresponding movies are given as supplemental material~\cite{supplemental}.}}
		\label{fig:controurmap_wakepattern}
	\end{figure}
	
	\begin{figure*}[!htbp]
		\centerline{\includegraphics[width = 1.0 \textwidth]{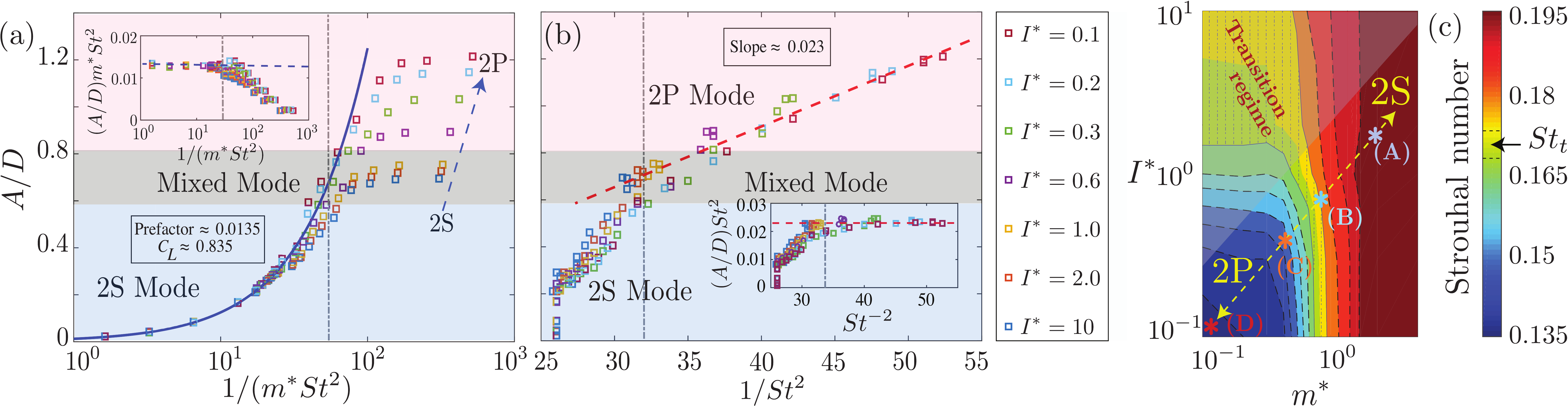}
		}
		\caption{(a)~Oscillation amplitude $A/D$ versus $1/({m}^*\ {St}^2)$  for buoyant and heavy cylinders. The dark blue curve shows the scaling relation ${A/D} \propto 1/({m}^*\ {St}^2$), which holds for ${A/D} \leq 0.6$. The inset to (a) shows the compensated plot; the plateau for small $1/(m^* {St}^2)$ demonstrates the robustness of this scaling. (b)~$A/D$ versus $1/ {St}^2$ for buoyant and heavy particles. For $A/D > 0.6$, the oscillation amplitude scales as ${A/D} \propto 1/{{St}^2}$, as also demonstrated by the plateau for large $1/{St}^2$ in the compensated plot in the inset to (b). (c)~Strouhal number contour map covering the [$m^*$- $I^*$] parameter space for the full family of buoyant and heavy cylinders from the present study. A transitional Strouhal number $St_t \approx 0.17$ separates the two scaling regimes shown in (a) and (b). 
		The dashed inclined line in (c) denotes the homogeneous cylinder cases, while the shaded region of high $I^*$ values is primarily of mathematical interest.}
		\label{fig:Scaling_Strouhal}
	\end{figure*}
	
	 The fluid motion governed by the incompressible Navier-Stokes equations may be written in the dimensionless form as:
\begin{equation}
\frac{\partial{\bf u}}{\partial{t}}  + \bf{u} . \nabla \bf {u}= \nabla \textit{p} +\frac{1}{\it{Re}} \nabla^2 \bf {u} + \bf f,
\label{model_equation}
\end{equation}
where {\bf u} is the velocity vector, \textit{p} is the pressure, $\it{Re}$ is the Reynolds number, and {\bf f} is the Eulerian body-force that is used to mimic the effects of the immersed body on the flow~\cite{zhu2014flow,zhu2014improved,zhu2014flexibility}. The direct numerical simulations provide an exact description of the flow-body interaction. However, for modeling purposes, 
the particle motion may be written in terms of the forces and moments exerted on the body. The Kelvin-Kirchhoff equations expressing linear and angular momentum conservation for the cylinder motion in a fluid may be extended to an incompressible flow containing vorticity~\cite{miloh1981generalization,mougin2001path}. The equations read:
\begin{eqnarray}
({m_p} + {m_a})\ [\frac{\text{d}{\bf U}}{\text{dt}} +  {\bf \Omega \times U}]   =  {\bf{F_{v}}} + (m_p - \rho_f \mathcal{V}_p) g ;
\label{forceeqn}
\end{eqnarray}
\begin{equation}
I\ \frac{\text{d}{\bf \Omega}}{\text{dt}} =  {\bf \Gamma_{v}};
\label{momenteqn}
\end{equation}
where $m_{p}$ is the particle mass, $m_a$ is the added mass, $I$  is the particle moment of inertia, $\bf{U}$ is the particle velocity vector, ${\bf \Omega}$ is the angular velocity, $\rho_f$ is the fluid density, $\mathcal{V}_p$ is the particle volume, and $g$ is the acceleration due to gravity. ${\bf F_v}$ and ${\bf \Gamma_v}$ represent the vortex forces and moments on the cylinder in a viscous fluid.

Eqs.~(\ref{forceeqn}) \& (\ref{momenteqn}) point to two parameter-dependences, namely the particle's mass and its moment-of-inertia. In addition, the rotation rate of the particle $\Omega$ is linked to $I$ and couples with eq.~(\ref{forceeqn}) through a force term. However, the dependence on $I$ was completely neglected for isotropic bodies despite the widespread variation in $I$ in almost all existing experimental studies~\cite{karamanev1996dynamics,horowitz2006dynamics,horowitz2008critical,horowitz2010effect,horowitz2010vortex,mathai2016translational,shafrir1965horizontal}. We therefore map out the [$m^*-I^*$] parameter space for rising and falling cylinders.


We begin with the case of a very buoyant cylinder~($m^* = 0.1$) rising in a quiescent fluid. \textcolor{black}{Fig.~\ref{fig:trajectory_separation_circulation}(a) shows the trajectory of the particle when its rotation is constrained~($\Omega =0$). The amplitude of oscillation $A/D \approx 0.65$. Next, we let the same particle auto-rotate while it rises through the fluid. Fig.~\ref{fig:trajectory_separation_circulation}(b)~shows the trajectory for $I^* = 0.1$. We observe an $85 \% $ enhancement in the oscillation amplitude due only to relaxing its contraint on rotation.} \textcolor{black}{Fig.~\ref{fig:trajectory_separation_circulation}(c) \& (d) show snapshots of the vorticity field in the vicinity of the particle for the two cases.
These observations indicate a clear link between the auto-rotation of the particle and the resulting translational and wake dynamics.}

{\color{black} Next, 
	we explore systematically the dynamics of heavy and buoyant cylinders, with $m^*$ and $I^*$ both varied in the range~[0.1;~10], i.e. covering very heavy~($m^*$ = 10) to very light~($m^*$ = 0.1) particles. It may be noted that in an actual experiment $m^*$ and $I^*$ are linked by the relation $I^* = m^* (R_{pG}^2 /R_{fG}^2)$, which imposes an upper bound $I^* \to 2 m^*$. Nevertheless, we vary the $I^*$  independently over a wide range~[0.1; 10] since it serves as an independent control parameter to modulate the particle's rotational freedom. The resulting translational and rotational amplitudes of vibration are shown in the contour maps in Fig.~\ref{fig:controurmap_wakepattern}(a)~\&~(b), respectively. The translational and rotational amplitudes depend on both $m^*$ and $I^*$. 
The change in wake-pattern as we move from upper-right to lower-left part of the contour map in Fig.~\ref{fig:controurmap_wakepattern}(a) \& (b) is shown through the sequence (A)$-$(D) in Fig.~\ref{fig:controurmap_wakepattern}(c). 
For large $m^*$, we always observe a \textit{2S}~(\textit{2-Single}) vortex mode \textbf{(A)}, which is identical to that of a cylinder fixed in a uniform flow. However, as we lower the $m^*$ and the $I^*$, we first observe a \textit{Mixed mode}~(\textbf{(B)} \& \textbf{(C)}), which gradually transitions to a clear \textit{2P}~(\textit{2-Pairs}) mode of wake vortices~\textbf{(D)}. From the figure, we see that the \textit{2S-2P} transition can occur for $m^*$ in the range [$0.2-1$], depending on the $I^*$. While a similar \textit{2S-2P} transition was also observed in the vortex-induced-vibrations literature~\cite{govardhan2002resonance}, the transition itself was thought be sudden at $m^* = 0.54$. Thus, unlike an elastically mounted or tethered particle, the extra rotational freedom present for a freely rising cylinder could be facilitating this gradual transition from \textit{2S} to \textit{2P} vortex mode. The same can be said about the oscillation amplitude, which grows gradually when $m^*$ and $I^*$ are reduced.}

Fig.~\ref{fig:controurmap_wakepattern} demonstrated the existence of regions where the mass did not play a serious role, along with others where the amplitude depended mainly on the mass. 
To model these, we adopt a simplified approach, where we assume that the vortex force on the cylinder~(from eq.~(\ref{forceeqn})) remains similar in intensity across the cases investigated. We describe this vortex force as $F_v \approx F_{v0} \sin \omega t$~\cite{bearman1984vortex,mathai2015wake}. Next, we assume that the system dynamics is governed by the vortex shedding frequency $\omega$.
For the freely rising case, the shedding frequency $\omega$ can get modified as we change $m^*$ and/or $I^*$. In the mass dominated regime, $\omega$ should primarily be a function of $m_p$. Thus, the transverse acceleration of the particle is given as $a_p \sim \frac{F_{v0}}{m_p} \sin \omega t$~\cite{mathai2015wake},
 which leads to a relation for the transverse motion:~$x_p \sim \frac{F_{v0}}{m_p \omega^2} \sin \omega t$. Non-dimensionalizing with the particle diameter $D$ and the mean rise velocity $U$, we obtain the dimensionless amplitude, ${A}/{D} \propto 1/(m^* \ {St}^2)$, where $m^*$ is the dimensionless mass and {\color{black}${St} \equiv \frac{\omega {D}}{2 \pi \ U}$ is the Strouhal number} or equivalently the dimensionless vortex shedding frequency~\cite{bearman1984vortex}. In Fig.~\ref{fig:Scaling_Strouhal}(a), we plot $A/D$ versus 1/($m^*$ \textit{St}$^2$) for the full range of parameters in the present study. The solid blue curve shows our prediction, where $F_v$ is estimated using the lift coefficient for flow past a fixed cylinder, i.e $C_L \approx 0.84$. The prediction matches well with the simulation results up to an oscillation amplitude $A/D \approx 0.5$. The inset to Fig.~\ref{fig:Scaling_Strouhal}(a) shows the compensated plot, where the plateau demonstrates the robustness of the scaling.

The prediction for the mass dominated regime is valid up to $A/D \approx$ 0.5. 
Beyond this, we observe branching of the oscillation amplitude for different values of the particle moment of inertia~(right half of Fig.~\ref{fig:Scaling_Strouhal}(a)). $A/D$ almost doubles in this regime~(0.6 to 1.2), solely due to a change in $I^*$. Therefore, we call this the moment of inertia dominated regime. Here the fluid added mass outweighs the particle mass. Therefore, the transverse oscillation amplitude in this regime may be expressed without an  $m^*$ dependence, as $A/D \propto~1/{St}^2$. $A/D$ versus $1/{St}^2$ for the full range of parameters is shown in Fig.~\ref{fig:Scaling_Strouhal}(b). $A/D$ scales linearly with $1/St^2$,  as demonstrated by the red dashed line. When compensated for this, we observe a plateau for $1/{St}^2 > 32$~(or equivalently, $St < 0.17$) in the inset. 
The scaling is valid for $A/D \geq$ 0.6, i.e. when the oscillation amplitudes are large. 


The above analysis suggests that the growth in oscillation amplitude is linked to the reduction in the dimensionless shedding frequency or $St$. 
Thus, a $St$ map~(as shown in Fig.~\ref{fig:Scaling_Strouhal}(c)) can summarise the dynamics of the entire family of heavy and light cylinders rising/falling through a quiescent fluid. 
\textcolor{black}{The clean \textit{2S} wake mode is found in the mass-dominated regime~(\textit{right-side}), and the clean \textit{2P} mode is restricted to the moment of inertia dominated regime~(\textit{lower-left}). The transition from \textit{2S} mode to \textit{2P} mode is not sudden, as was found for elastically mounted particles~\cite{govardhan2004critical,govardhan2005vortex}, but gradual and marked by the appearance of a $\textit{Mixed}$ wake mode in the $0.5 < A/D < 0.7$ range.} 

\begin{figure}[!htbp]
	\centerline{\includegraphics[width = 0.5 \textwidth]{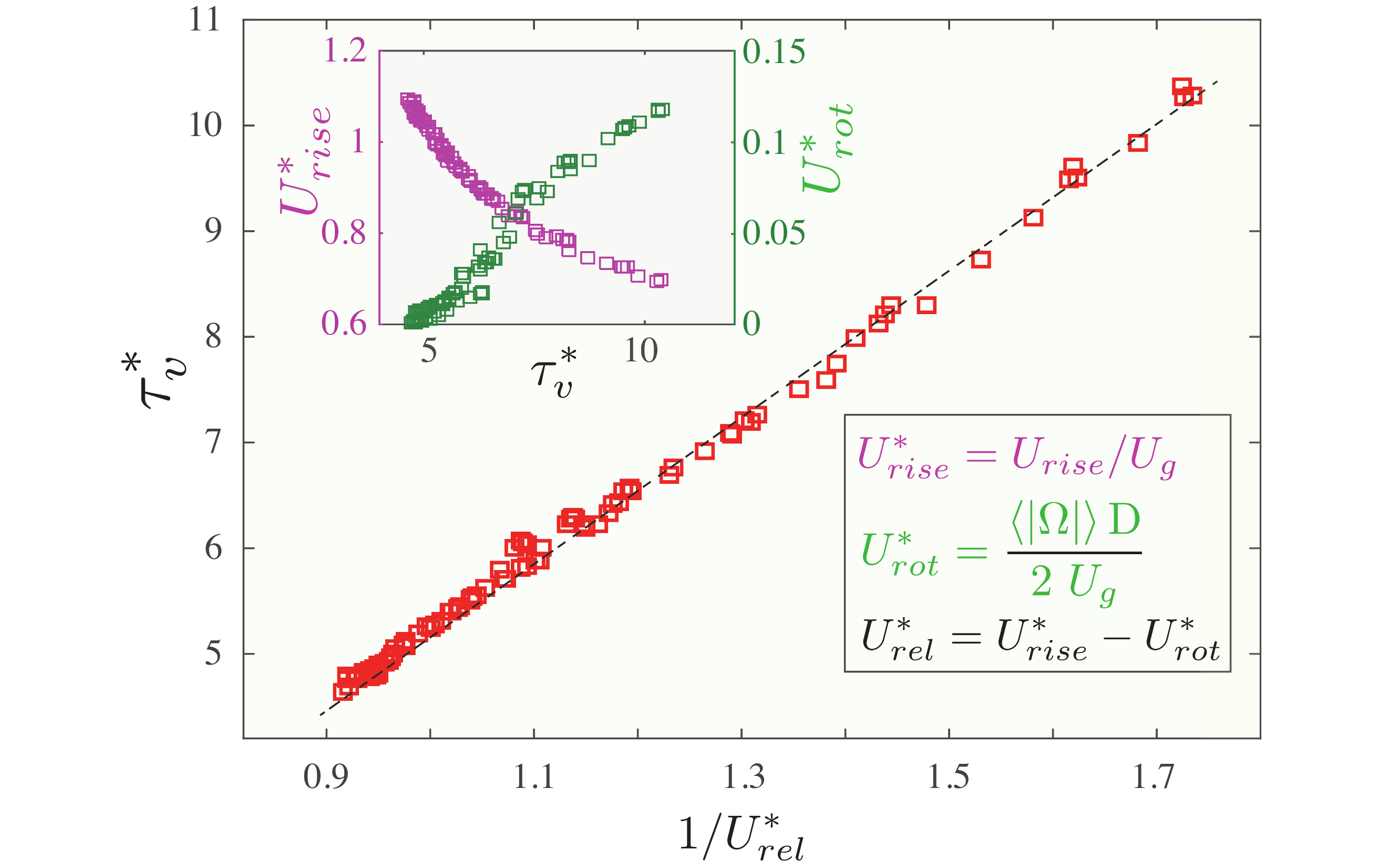}
	}
	\caption{Vortex-shedding time $\tau_v^*$ vs the inverse of the relative speed $1/U_{rel}^*$ for all heavy and buoyant cylinders from the present study. The dashed line suggests a nearly linear dependence. The inset shows the mean rise speed~$U_{rise}^*$ and the mean rotational speed~$U_{rot}^* \equiv  \frac{\left < |\Omega| \right > {D}}{2 \ U_g}$ vs $\tau_v^*$, where $\Omega$ is the instantaneous angular velocity of the cylinder, $D$ is the cylinder diameter, and $U_g \equiv \sqrt{g {D} |1 - \rho_p/\rho_f|}$ is a gravitational velocity scale~\cite{mathai2015wake}. 
	}
	\label{fig:wakesimilarity}
\end{figure}

\textcolor{black}{The main conclusions of the present study are valid even at a lower $Ga$~($=220$). At this $Ga$~(or corresponding $Re$), the flow can be considered predominantly two-dimensional, as suggested by Leontini \etal~\cite{leontini2007three}. Here again, we observe two distinct scaling regimes: a mass-dominated regime for $0 < A/D < 0.3$, and a moment of inertia dominated regime for $0.4 < A/D < 0.75$, with their respective $1/(m^* St^2)$ and $1/St^2$ scalings. At the same time, the maximum oscillation amplitude is lower ($A/D \approx 0.75$), and the wake transition is less distinct, owing to the greater viscous effects. In future work, we will provide a systematic account of the $Ga$ (or $Re$) effects on the dynamics and wakes of rising/falling cylinders.}


 
 The simplified model adopted here has enabled us to understand the motions and wakes of heavy and light cylinders rising/falling through still fluid. From the assumption that the vortex forces remain similar in intensity, but change only in the duration of their action, we can explain the growth in the oscillation amplitude. Since the Strouhal number represents the dimensionless frequency of vortex shedding, the fact that it is reduced
means that the vortex shedding is retarded~\cite{namkoong2008numerical}. While the precise mechanism by which rotation induces this is not clear, one can qualitatively explain the behavior.
 \textcolor{black}{For this, we compare the rotational response of a high moment of inertia cylinder with that of a low moment of inertia cylinder~(Fig.~\ref{fig:trajectory_separation_circulation}(c) \& (d)). The high moment of inertia cylinder resists rotation, while the low moment of inertia cylinder yields to the fluid torques and rotates. On the side where vorticity is shed~(see Fig.~\ref{fig:trajectory_separation_circulation}(c) \& (d)), the particle's rotation is along the mean flow direction.} Therefore, the relative speed at the cylinder surface ${U}_{{rel}}= (U_{rise} - U_{rot})$ is reduced for the low moment of inertia cylinder, which could cause the developing shear-layer to remain attached for longer. Thus the vortex shedding time $\tau_v \propto 1/{U}_{rel}$ for the low moment of inertia cylinder, as compared to $\tau_v \propto 1/{U_{rise}}$ for the non-rotating cylinder. This corresponds to a reduction in the dimensionless frequency for the low moment of inertia cylinder. In Fig.~\ref{fig:wakesimilarity}, we find evidence that the dimensionless vortex shedding timescale $\tau_v^* \equiv \frac{\tau_v \ U_g}{D}$ increases almost linearly with the dimensionless inverse relative speed~$1/U_{rel}^* \equiv \frac{U_g}{U_{rel}}$, where $U_g = \sqrt{g D |1-\rho_p/\rho_f|}$ is a gravitational velocity scale~\cite{mathai2015wake}.

 \textcolor{black}{In summary, the present study has demonstrated that the path oscillations of a buoyant particle can be linked to its rotational freedom, as speculated by Ryskin \& Leal in 1984~\cite{ryskin1984numerical}. Our findings remain to be experimentally confirmed using cylinder-rising experiments, such as those conducted in the past~\cite{horowitz2006dynamics}, but with the rotational inerta as an additional control parameter.
 The insights gained here could be extended to isotropic bodies in 3D, such as buoyant spherical particles~\cite{kelley1997path,wu2002experimental,mougin2001path,ern2012wake}. In an ongoing work, we have noticed  the validity of this for buoyant spheres rising in still fluid, as well as for buoyant spheres rising through a turbulent flow. These will be the focus of a future work.}



Varghese Mathai and  Xiaojue Zhu contributed equally to this work, and their names in the author-list are interchangeable.
We thank Guowei He and Xing Zhang for the initial development of the code used here.
We gratefully acknowledge J.~Magnaudet, L.~van~Wijngaarden, S.~Wildeman, and A.~Prosperetti for useful discussions. This work was financially supported by the STW foundation of the Netherlands, and the Foundation for Fundamental Research on Matter~(FOM), which is a part of the Netherlands Organisation for Scientific Research~(NWO). CS acknowledges the financial support from Natural Science Foundation of China under Grant No. 11672156.
We thank the Dutch Supercomputing Consortium SurfSARA, the Italian supercomputer FERMI-CINECA through the PRACE Project No. 2015133124 and the ARCHER UK National Supercomputing Service through the DECI Project 13DECI0246 for the allocation of computing time. 
}


\end{document}